# Effects of Pressure on the Electronic and Structural Properties of LaOFeAs

Yong Yang and Xiao Hu

WPI Center for Materials Nanoarchitectonics (MANA)

National Institute for Materials Science, Tsukuba 305-0044, Japan

We studied the pressure effects on the electronic and structural properties of LaOFeAs by first-principles calculations. For the anti-ferromagnetic (AFM) phase with stripe-like aligned Fe spins, the electronic density of states at the Fermi level ($N$ ($E_F$)) slightly descends first with increasing applied pressure, then bounces up with further increasing pressure (or decreasing volume), and reaches its maximum at ~ 29.2 GPa with the volume ~ 80% of the ambient pressure value ($V_0$). Similar behavior is expected in the physical quantities that are proportional to $N$ ($E_F$), such as the electronic specific heat and the Pauli paramagnetic susceptibility. At this volume ($V = 0.8V_0$), the LaOFeAs crystal undergoes a structural phase transition from the orthorhombic structure to the tetragonal one, which is accompanied by the disappearance of the long-ranged AFM order.

# I. INTRODUCTION

The discovery of superconductivity in the iron-based materials has generated great excitements in scientific communities. Soon after the reported superconductivity in $LaO_{1-x}F_xFeAs$ with a transition temperature ($T_c$) of ~ 26 K [1], a number of analogues have been synthesized, and $T_c$ was soon boosted to ~ 55 K [2, 3]. The newly discovered iron-based superconducting materials are generally named as "Fe-pnictides". So far, four categories of iron-based superconductors have been discovered: the 1111 series (ROFeAs, R = rare earth elements; AeFFeAs, Ae = alkaline earth metals elements) [1-5], the 122 series ($AeFe_2As_2$; $AFe_2As_2$, A = alkali metals elements) [6-8], the 111 series (AFeAs) [9], and the 11 series (FeSe) [10]. These four series of Fe-pnictides are found to exhibit some common features in both bonding geometries and electronic structures [11-15].

To achieve superconductivity, two methods are commonly used — doping and applying pressure. Either electron doping or hole doping can introduce superconductivity [1-8]. For some iron-based materials (e.g., $LaO_{1-x}F_xFeAs$, $FeSe_x$) that are superconductors at ambient pressure, their superconducting temperatures $T_c$ are observed to increase in a certain region of applied pressure (several GPa) [16-18]. Moreover, the undoped LaOFeAs and $AFe_2As_2$ (A = Ca, Sr, Ba), non-superconducting materials at ambient pressure, are observed to turn into superconductors at appropriate high pressures [19-22]. More details on the pressure effects on the Fe-pnictides superconductors can be found in a recent review article [23]. In spite of the rapid progress in the accumulation of experimental data, little is understood on the microscopic mechanism of the pressure-induced superconductivity, especially from the level of electronic structures. Theoretical efforts, especially first-principles calculations, have been helpful to elucidate the electronic and magnetic properties of the Fe-pnictides [11-15, 24-27]. Compared to the numerous works at ambient pressure, theoretical investigations on the pressure effects on the Fe-pnictides are relatively limited except several ones using first-principles calculations [26-29]. Up to now, there is a lack of systematic investigations on the pressure response of the electronic density of states at the Fermi level, which plays a key role in the phenomena of superconductivity. More efforts from the theoretical side are necessary.

In this work, we study the pressure effects on the electronic and structural properties of the undoped LaOFeAs crystal using first-principles calculations. The configuration with in-plane stripe-like anti-ferromagnetic (AFM) Fe spins is studied here, which is the magnetic ground state at low temperature suggested by experiments [30]. Our calculations show that, the electronic density of states at the Fermi level, $N(E_F)$, has a tendency of slightly dropping with decreasing crystal volume (ascending pressure) first, then surprisingly jumps up and arrives at a maximum, about 8 times larger than the value at ambient pressure. $N(E_F)$ then decreases with further increasing applied pressure. The enhancement of $N(E_F)$ is in accordance with our calculated electronic band structures for descending crystal volumes. The LaOFeAs crystal undergoes a structural phase transition from the orthorhombic structure (*Cmma* space group) to tetragonal structure (*P4/nmm* space group) at around 29.2 GPa, and the magnetic moment of the Fe atoms disappears at this structural transition.

## II. COMPUTATIONAL DETAILS

The calculations are performed by the Vienna *ab initio* simulation package (VASP) [31, 32], which is based on density functional theory (DFT), using a plane wave basis set and the PAW potentials [33, 34]. The exchange-correlation interactions are described by the generalized gradient approximation (GGA) with PBE type functional [35]. The energy cutoff for plane waves is 600 eV. If not specially noted, a 12×12×6 Monkhorst-Pack k-points grid [36] is used in the calculations. The atomic and magnetic structure of LaOFeAs considered in this work is shown in Fig. 1. This is a $\sqrt{2}\times\sqrt{2}\times1$ supercell with stripe-like spin arrangement in the FeAs plane. It is one of the mostly used models for studying the electronic and magnetic properties of the LaOFeAs [12, 25-27, 29]. Such an AFM plane has two possible arrangements along the *c*-axis: the ferromagnetic (FM) stacking and the AFM stacking. From our calculations, the difference of total energy between the FM and AFM stacking along *c*-axis is less than 1 meV/super-cell, which is within the precision limit. This indicates that the magnetic interactions between two neighboring FeAs planes are very weak, and justifies the validity of using a $\sqrt{2}\times\sqrt{2}\times1$ cell in *ab initio* simulations [27].

## III. RESULTS AND DISCUSSION

Starting from the experimentally determined tetragonal structure at 175 K [30], we did structural optimization on the atomic positions and the cell axes with the cell volume fixed to the experimental value. From our calculation, the difference between the optimized volume and the experimental volume at ambient pressure is less than 1%. Table I summarizes the cell parameters of LaOFeAs obtained by our calculations.

The ground state geometries compare well with experiments at low temperature [30], for both the length of cell axes and the internal distances and angles. An orthorhombic cell is obtained and the angle between the two lattice vectors in FeAs plane is ~ 90.67º, very close to the experimental value of 90.28º [30]. Compared to the four different As-Fe-As angles provided in the Table 2 of Ref. [30], only three different As-Fe-As angles are found in our calculations. This is due to the fact that there are totally six As-Fe-As angles in a tetrahedrally bonded structure, and they are two-fold degenerate along each of the three lattice vectors. This degeneration is preserved in our DFT calculations. In the case of tetragonal lattice, the *a* & *b* directions are also degenerate, which leads to only two different angles values of As-Fe-As, as is evidenced in the Table 1 of Ref. [30]. For the orthorhombic case, the two values of As-Fe-As angle (114.2(5)º vs 113.8(4)º) from neutron diffraction experiments are within the error bars [30], which implies that the two angles may take the same value in more precise measurements at lower temperature. Similar situation is found in the case of La-O-La angles, on which no experimental data are available yet.

The calculated magnetic moment of Fe atom is 1.97 $\mu_B$, with $\mu_B$ being the Bohr magneton. This value is much larger than the value ~ 0.36 $\mu_B$ measured by neutron diffraction [30]. Controversial results are obtained from DFT calculations using optimized crystal structures: the magnetic moment of Fe is reported to ~ 0.48 $\mu_B$ in Ref. [25], while a value of ~ 0.75 $\mu_B$ in Ref. [27], and a value of ~ 2 $\mu_B$ in a recent work [29], which is close to our results. Clearly more efforts are needed to resolve this problem.

To study the pressure effects, we decreased the volume of LaOFeAs crystal gradually. For each fixed volume, we relaxed the shape of the super-cell and the positions of atoms and then calculate the electronic density of states (DOS) at the Fermi level (referred as $N(E_F)$ hereafter). Figure 2(a) shows the variation of $N(E_F)$ as a function of the volume of LaOFeAs. With the increasing of applied pressure, or the contraction of volume, $N(E_F)$ decreases first and reaches its minimum when the volume is ~ 90% of the equilibrium volume at ambient pressure (denoted by $V_0$). Then $N(E_F)$ bounces up with further increasing pressure, and arrives at its maximum at ~ $0.8V_0$, at which $N(E_F)$ is about 8 times larger than the value at volume $V = V_0$. $N(E_F)$ starts to drop gradually with even higher pressure and larger volume contraction. The DFT calculations are performed with high accuracies for the value of $N(E_F)$, with the k-point meshes in the Brillouin zone increasing to 16×16×10 and 16×16×12. As displayed in Fig. 2(a), $N(E_F)$ already shows good convergence as indicated by the results obtained by the two sets of k-meshes.

We then calculated the pressure associated with the volume contraction, which is displayed in Fig. 2(b). The pressure is evaluated as the first derivative of total energy with respective to volume. The pressure at $V = 0.8V_0$ is calculated to be ~ 29.2 GPa. This implies that the electronic specific heat ($\gamma \propto N(E_F)$), and the Pauli susceptibility ($\chi \propto N(E_F)$) of this system at applied pressure of ~ 29.2 GPa will be about 8 times larger than the value measured at ambient pressure.

We have further analyzed the contributions of each type of atoms to the electronic DOS around the Fermi level, as shown in Fig. 3. For both $V = V_0$ and $V =$ $0.8V_0$, the major part of the DOS around the Fermi level comes from the Fe atoms. From Fig. 3(c), it is clear that the great enhancement of $N(E_F)$ is mainly contributed from the 3d orbitals of Fe atoms.

Figure 4 shows the calculated electronic band structure of LaOFeAs at several volumes ($V = V_0$, $0.9V_0$, $0.8V_0$, $0.7V_0$). The bands are calculated along some high-symmetry lines in the Brillouin zone. It is clear that more than one bands run across the Fermi level, and thus the metallic and multiband features are kept under pressure. Generally, the more crossing band points at the Fermi level, the higher $N(E_F)$ will be expected. Most crossing points are found in the case of $V = 0.8V_0$, then followed by $V = 0.7V_0$, $V_0$ and $0.9V_0$ in order. This trend is consistent with the calculated $N(E_F)$ as shown in Fig. 2(a). The band structures obtained in the present work at $V = V_0$ are comparable with previous DFT all-electron full potential calculations [24].

Variations of geometry parameters of the primitive cell of LaOFeAs crystal and magnetic moment of the Fe atoms upon pressure application are displayed in Figs. 5(a) ~ (c). In Fig. 5(a), the ratio between ($c/c_0$) and ($a/a_0$) keeps descending with decreasing volume, which agrees with previous experimental and theoretical studies [27, 29, 37], and indicates that the LaOFeAs crystal is more compressible along *c* axis than along *a* (or *b*) axis. This can be understood from the layered crystal structure: the interlayer coupling is weaker than the bonding in the same LaO and FeAs layers.

Figure 5(b) shows the evolution of the angle between the *a* and *b* axes of the primitive cell. The angle decreases with the contraction of crystal volume, and reaches 90º at V ~ 0.8$V_0$ and then remains constant with further contraction of the volume. This means that the LaOFeAs crystal undergoes a structural phase transition from the orthorhombic structure at ambient pressure to the tetragonal structure at the applied pressure of ~ 29.2 GPa and beyond. Table II gives the super-cell structural parameters calculated at V = 0.8$V_0$. The angle between the *a* and *b* axes is 89.98º, very close to 90º. In addition, only very small differences are found in the two Fe-As bond lengths, and the two As-Fe-As angles (105.05º vs 105.08º) and La-O-La angles (105.87º vs 105.90º). Within the computational accuracy, the angle 89.98º can be taken as 90º, and the other parameters under comparison can be safely regarded as the same value, and thus demonstrates the degeneration of *a* and *b* axes. These cell parameters provide direct evidences for the structural phase transition under pressure.

At the same time, the calculated magnetic moment of Fe decreases monotonously with reducing volume and disappears at V = 0.8$V_0$, as shown in Fig. 5(c). Similar results on the Fe magnetic moment were reported in two earlier works [27, 29]. On the other hand, the difference between our work and the results presented in Ref. [27] and Ref. [29] is distinguished: In addition to the great enhancement of *N* ($E_F$) under pressure (Fig. 2), we predict for the first time the pressure-induced structural phase transition at V = 0.8$V_0$ or at pressure of ~ 29.2 GPa.

The results discussed in Fig. 5 show again that the structural phase transition in the crystal of undoped Fe-pnictides is usually accompanied by the appearance/disappearance of the long-range spin density wave (SDW) order, as demonstrated by experiments in the 1111 and 122 systems by changing the temperature [30, 38].

We note here that, as a well-known fact, even the state-of-the-art DFT calculations cannot give the correct superconducting electronic states. Thus, the predictions regarding electronic and magnetic properties don't apply to superconducting region. As reported in Ref. [19], pure LaOFeAs crystal becomes superconductor under pressure of ~ 12 GPa, with a superconducting temperature of ~ 20 K. This implies the disappearance of AFM order at temperatures lower than 20 K when applied pressure > 12 GPa. When a competition between the magnetic order and superconductivity takes place, DFT calculations overestimate the territory of magnetic order. The disagreement between the DFT result on the pressure at which the magnetic order disappears, ~ 29.2 GPa, and the experimental result at ~ 12 GPa is understandable.

In light of this notice, we consider that the above electronic, magnetic and crystalline structural evolution can be tested by experiments at temperatures $T$ above the superconducting transition temperature $T_c$ but lower than Nèel temperature $T_N$, i.e., $T_c < T < T_N$. In such a temperature region, the LaOFeAs keeps its AFM ordering but no superconducting states are involved in the measurements. In the Fe-pnictides, superconductivity usually appears when the anti-ferromagnetism or SDW is suppressed. Therefore, the results discussed above may hint that superconductivity will be enhanced by high pressure.

## IV. CONCLUSIONS

In conclusion, the pressure effects on the electronic and structural properties of the LaOFeAs crystal have been studied by first-principles calculations. The electronic density of states at the Fermi level ($N(E_F)$) shows slight descending with decreasing crystal volume, then jumps up with further contraction of the volume and arrives at its maximum at $V = 0.8V_0$, corresponding to an applied pressure of ~ 29.2 GPa. The maximum value of $N(E_F)$ is about 8 times larger than the value at ambient pressure. The major part of $N(E_F)$ comes from the 3d orbitals of Fe atoms. To the best of our knowledge, such behavior of $N(E_F)$ is reported for the first time by present work. Thus great enhancement in the electronic specific heat and the Pauli magnetic susceptibility of undoped LaOFeAs are expected at the pressure $P \sim 29.2$ GPa. At this pressure, our calculations predict a structural phase transition from the original orthorhombic structure to the tetragonal structure, which is accompanied with magnetic transition from the stripe-like anti-ferromagnetic state to paramagnetic state. This behavior is similar to that observed upon temperature variation [30]. We hope that the results obtained by our calculations can be verified by future experiments.

## ACKNOWLEDGEMENTS

Calculations were performed on SR11000 (HITACHI) in NIMS. This work was supported by WPI Initiative on Materials Nanoarchitectonics, MEXT, Japan.

**References:**

[1] Y. Kamihara, *et al.*, J. Am. Chem. Soc. **130**, 3296 (2008).

[2] X. H. Chen, *et al.*, Nature **453**, 761 (2008).

[3] Z. A. Ren, *et al.*, Europhys. Lett. **83**, 17002 (2008).

[4] S. Matsuishi, *et al.*, J. Phys. Soc. Jpn. **77**, 113709 (2008).

[5] F. Han, *et al.*, Phys. Rev. B **78**, 180503(R) (2008).

[6] M. Rotter, *et al.*, Phys. Rev. Lett. **101**, 107006 (2008).

[7] K. Sasmal, *et al.*, Phys. Rev. Lett. **101**, 107007 (2008).

[8] G. Wu, *et al.*, J. Phys.: Condensed Matter **20**, 422201 (2008).

[9] X. C. Wang, *et al.*, Solid State Comm. **148**, 538 (2008).

[10] F. C. Hsu, *et al.*, Proc. Nat. Acad. Sci. **105**, 14262 (2008).

[11] D. J. Singh and M.-H. Du, Phys. Rev. Lett. **100**, 237003 (2008).

[12] G. Xu, *et al.*, Europhys. Lett. **82**, 67002 (2008).

[13] C. Liu, *et al.*, Phys. Rev. Lett. **101**, 177005 (2008).

[14] D. J. Singh, Phys. Rev. B **78**, 094511 (2008).

[15] A. Subedi, *et al.*, Phys. Rev. B **78**, 134514 (2008).

[16] H. Takahashi, *et al.*, Nature **453**, 376 (2008).

[17] W. Lu, *et al.*, New J. Phys. **10**, 063026 (2008).

[18] Y. Mizuguchi, *et al.*, Appl. Phys. Lett. **93**, 152505 (2008).

[19] H. Okada, *et al.*, J. Phys. Soc. Jpn. **77**, 113712 (2008).

[20] M. S. Torikachvili, *et al.*, Phys. Rev. Lett. **101**, 057006 (2008).

[21] T. Park, *et al.*, J. Phys.: Condens. Matter **20**, 322204 (2008).

[22] P. L. Alireza, *et al.*, J. Phys.: Condens. Matter **21**, 012208 (2009).

[23] C. W. Chu and B. Lorenz, arXiv:0902.0809v1 (unpublished).

[24] Z. P. Yin, *et al.*, Phys. Rev. Lett. **101**, 047001 (2008).

[25] T. Yildirim, Phys. Rev. Lett. **101**, 057010 (2008).

[26] T. Yildirim, Phys. Rev. Lett. **102**, 037003 (2009).

[27] I. Opahle, *et al.*, Phys. Rev. B **79**, 024509 (2009).

[28] W. Xie, *et al.*, Phys. Rev. B **79**, 115128 (2009).

[29] S. Lebègue, Z. P. Yin, and W. E. Pickett, New J. Phys. **11**, 025004 (2009).

[30] C. de la Cruz, *et al.*, Nature **453**, 899 (2008).

[31] G. Kresse and J. Hafner, Phys. Rev. B **47**, 558 (1993).

[32] G. Kresse and J. Furthmüller, Phys. Rev. B **54**, 11169 (1996).

[33] P. E. Blöchl, Phys. Rev. B **50**, 17953 (1994).

[34] G. Kresse and D. Joubert, Phys. Rev. B **59**, 1758 (1999).

[35] J. P. Perdew, *et al.*, Phys. Rev. Lett **77**, 3865 (1996).

[36] H. J. Monkhorst and J. D. Pack, Phys. Rev. B **13**, 5188 (1976).

[37] G. Garbarino, *et al.*, Phys. Rev. B **78**, 100507(R) (2008).

[38] J. W. Lynn and P. Dai, Physica C **469**, 469 (2009), and references therein.

Table I. Optimized structural parameters of LaOFeAs crystal at ambient pressure, with stripe-like anti-ferromagnetic arrangement of Fe spin. Here the data are for the primitive cell, and γ denotes the angle between the lattice vectors (along *a* and *b* axes) in the FeAs plane.

$V = V_0 = 141.9$ Å$^3$, *a* = 4.04 Å; *b* = 4.04 Å; *c* = 8.68 Å; γ = 90.67°.

| Distances (Å) | this work | Expt.[a] | Angles (°) | this work | Expt.[a] |
|---|---|---|---|---|---|
| La-O (×4) | 2.37 | 2.394, 2.342 | As-Fe-As | 115.80° | 114.2(5) , 113.8 (4) |
| Fe-As (×4) | 2.39 | 2.398, 2.405 | As-Fe-As | 106.91° | 107.47(6) |
| La-As (×2) | 3.38 | 3.369 | As-Fe-As | 105.90° | 107.06(6) |
| La-As (×2) | 3.41 | 3.380 | La-O-La | 116.99° | --- |
| Fe-Fe | 2.84 | 2.8409 | La-O-La | 106.35° | --- |
| Fe-Fe | 2.88 | 2.8548 | La-O-La | 105.34° | --- |

[a]Ref. [30].

Table II. Similar as Table I but for volume $V = 0.8 V_0$.

$V = 0.8V_0 = 113.52$ Å$^3$, a = 3.88 Å; b = 3.88 Å; c = 7.54 Å; γ = 89.98°.

| | | | |
|---|---|---|---|
| La-O (×4) | 2.28 Å | As-Fe-As | 118.70° |
| Fe-As (×4) | 2.26 Å | As-Fe-As | 105.05° |
| La-As (×2) | 3.09 Å | As-Fe-As | 105.08° |
| La-As (×2) | 3.09 Å | La-O-La | 116.91° |
| Fe-Fe | 2.75 Å | La-O-La | 105.87° |
| Fe-Fe | 2.74 Å | La-O-La | 105.90° |

**Figure Captions:**

FIG. 1 (color online) Schematic diagram for the $(\sqrt{2} \times \sqrt{2})$ super-cell of LaOFeAs crystal with stripe-like anti-ferromagnetic spin arrangement of Fe atoms.

FIG. 2 (color online) **(a)** Electronic density of states at the Fermi level as a function of volume ratio $V/V_0$, with V being the volume of LaOFeAs and $V_0$ the experimental volume of LaOFeAs at ambient pressure. The calculations are performed with two different k-points grids: 16×16×10, and 16×16×12. **(b)** Calculated pressure at the descending volumes of with an interval of $0.05V_0$.

FIG. 3 (color online) From top to bottom panels: the electronic density of states (DOS) contributed from La, O, Fe, As atoms at $V = V_0$ and $V = 0.8V_0$, and the DOS projected on the 3d orbitals of Fe atoms at $V = V_0$ and $V = 0.8V_0$, respectively. The Fermi level is indicated by vertical dash line.

FIG. 4 (color online) Band structures of LaOFeAs at four descending volumes with an interval of $0.1V_0$: **(a)** $V = V_0$; **(b)** $V = 0.9V_0$; **(c)** $V = 0.8V_0$; **(d)** $V = 0.7V_0$. The coordinates for the symmetry points in the reciprocal space are as follows: S = (0.5, 0.5, 0), R = (0.5, 0.5, 0.5), Γ = (0, 0, 0), X = (0.5, 0, 0), Y = (0, 0.5, 0), Z = (0, 0, 0.5).

**FIG. 5** (color online) **(a)** c/a ratio (normalized to ambient pressure value), **(b)** angle between the two lattice vectors in the FeAs plane of the primitive cell, and **(c)** magnetic moment of Fe atom, as a function of $V/V_0$.

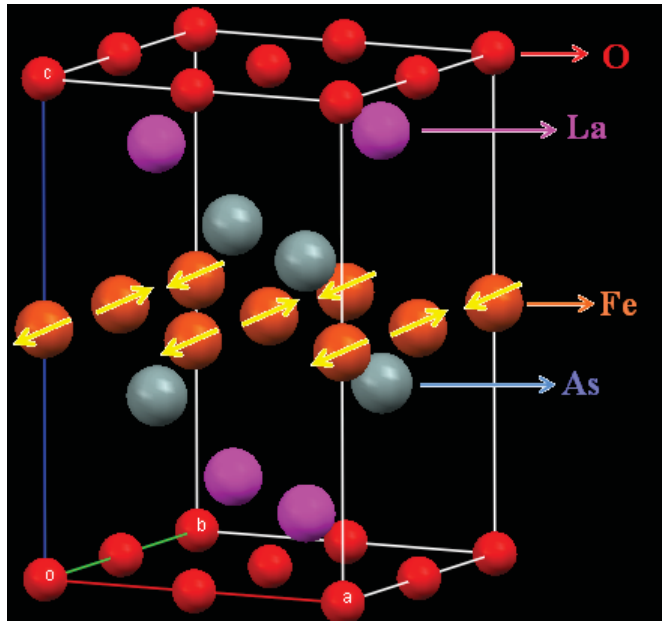

**FIG. 1 (of 5) Yang & Hu**

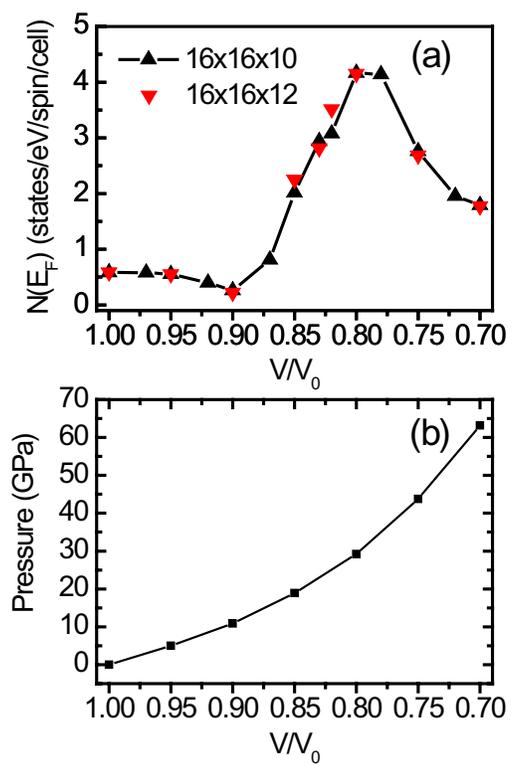

**FIG. 2 (of 5) Yang & Hu**

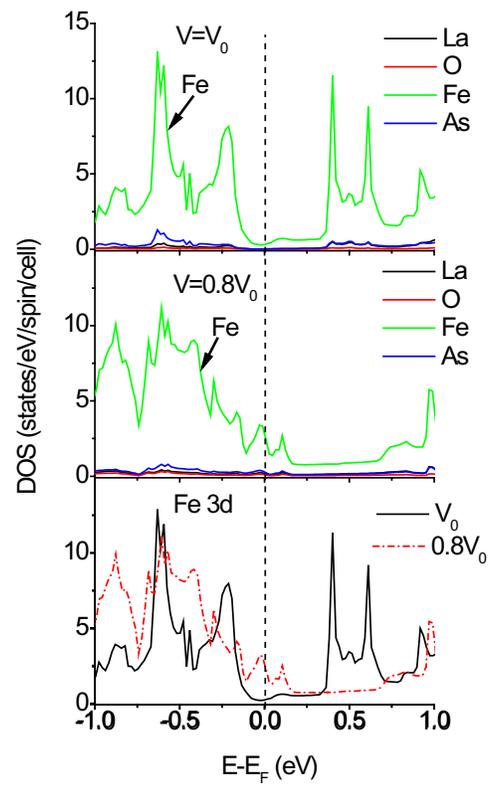

**FIG. 3 (of 5) Yang & Hu**

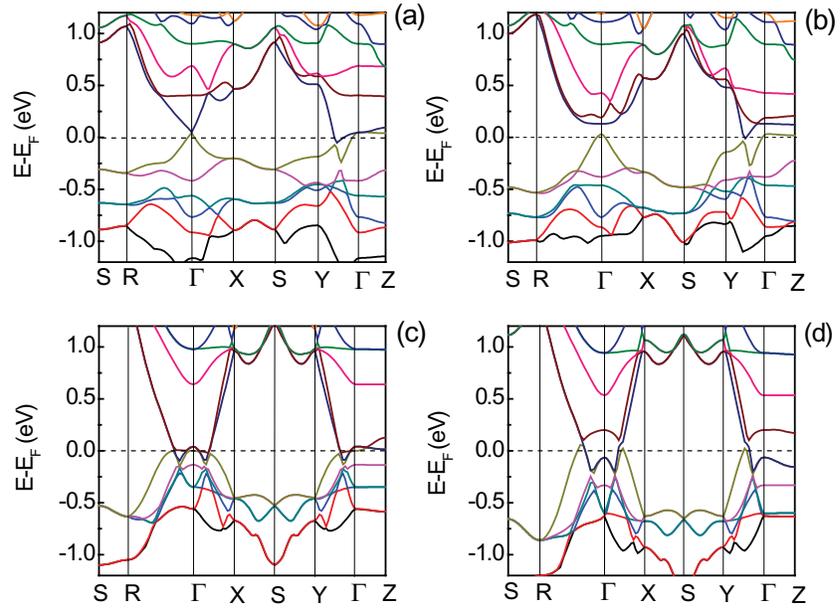

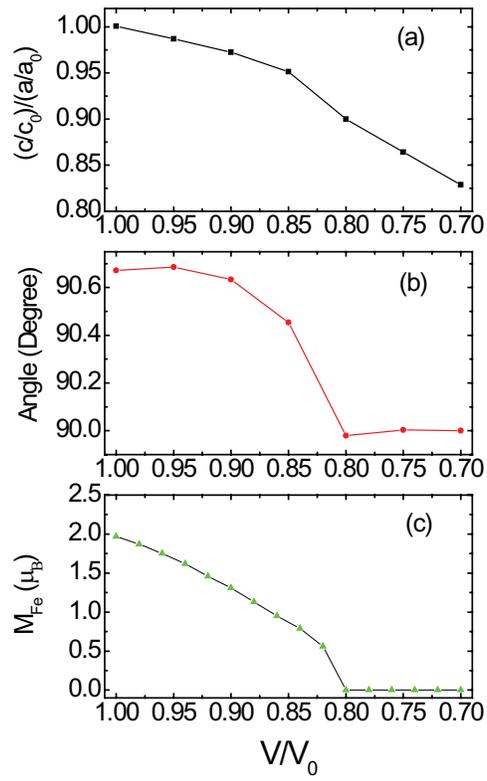

**FIG. 5 (of 5) Yang & Hu**